\documentclass[12pt]{article}
\usepackage{amssymb}

\begin{document}

\newcommand{\beq}{\begin{equation}}
\newcommand{\eeq}{\end{equation}}
\newcommand{\be}{\begin{eqnarray}}
\newcommand{\ee}{\end{eqnarray}}
\renewcommand{\vec}[1]{{\bf #1}}
\newcommand{\vecg}[1]{\mbox{\boldmath $#1$}}
\newcommand{\grpicture}[1]
{
    \begin{center}
        \epsfxsize=200pt
        \epsfysize=0pt
        \vspace{-5mm}
        \parbox{\epsfxsize}{\epsffile{#1.eps}}
        \vspace{5mm}
    \end{center}
}

\date{SUBATECH--2003/03 \\
ITEP-TH--05/03}

\title{Effective
Lagrangian for 3d ${\cal N} = 4$ SYM
theories for any gauge group and  monopole moduli spaces}

\author{K.G. Selivanov \\
{\normalsize ITEP, B.Cheremushkinskaya 25, Moscow 117218, Russia}\\
\and\\
A.V. Smilga \\
{\normalsize SUBATECH, Universit\'e de
Nantes,  4 rue Alfred Kastler, BP 20722} \\ {\normalsize Nantes  44307, France. }
\footnote{on leave of absence from ITEP, Moscow, Russia.}}

\maketitle

\begin{abstract}
We construct  low energy effective Lagrangians for 3d ${\cal N}=4$
supersymmetric
Yang-Mills theory with any gauge group.
They represent supersymmetric $\sigma$ models at
hyper--K\"ahlerian manifolds of dimension $4r$ ($r$ is the rang
of the group).
In the asymptotic region, perturbatively exact explicit
expression for the metric are written.
We establish the relationship of this metric
with  the TAUB-NUT metric describing
the perturbatively exact effective
Lagrangians for unitary groups and  monopole moduli spaces:
the former is obtained out of the latter by a proper
hyper--K\"ahlerian reduction.
We describe in details the reduction procedure for $SO/Sp/G_2$
gauge groups, where it can also be given a natural interpretation
in $D$-brane language.
We conjecture that the exact nonperturbative
metrics can be obtained by a similar hyper--K\"ahlerian
reduction from the corresponding multidimensional Atiyah--Hitchin metrics.

\end{abstract}

\section{Introduction}

In a very well-known  paper \cite{WS1},
Seiberg and Witten have found the exact effective
low--energy Lagrangian for 4d ${\cal N} =2$ supersymmetric
Yang--Mills theories for the $SU(2)$ gauge group. This result was
later generalised to other gauge groups  \cite{reviews,Nekr}.
A related interesting question is what is the form of the effective
Lagrangian in lower--dimensional "descendants" of 4d ${\cal N} =2$
SYM theory, the models obtained from it by dimensional reduction.

Going down to 3d, we obtain  ${\cal N} =4$ (in the
3--dimensional sense)
SYM theory. Its effective Lagrangian was constructed  in the
$SU(2)$ case in \cite{Seiberg1,WS2}. It represents a
hyper--K\"ahlerian supersymmetric $\sigma$ model on the
{\it Atiyah--Hitchin} manifold \cite{AH}. This metric also
describes the dynamics of two interacting BPS
monopoles \cite{GM}, the moduli space of classical vacua in SYM theory
exactly coinciding with the monopole moduli space.
Indeed, the low--energy dynamics in the former involves
3 scalars originating from the components of
Abelian vector potentials in the original
6d ${\cal N}=1$ theory in the reduced dimensions and a scalar
dual to the 3d photon. And this corresponds to a relative distance
of two monopoles and their relative phase.

In the asymptotics when the distance $|\vec{r}|$ between the
monopoles becomes large, this metric goes
over to the TAUB-NUT metric (with negative mass term),
  \be
\label{TAUBNUT}
2ds^2 \ =\  \left( 1 - \frac {g^2}{2\pi |\vec{r}|}
\right) d\vec{r}^2 + \frac {
\left(d\Psi + \frac {g^2}{2\pi} \omega  \right)^2}
{\left( 1 - \frac {g^2}{2\pi |\vec{r}|}
\right)} \ ,
 \ee
where $\Psi$ is the relative phase of the monopoles and
 \be
 \label{Dirac}
\omega  ( \vec{r})\  = \  \cos \theta \ d\phi
  \ee
  is the Abelian connection describing a Dirac monopole. The factor 2 on the left side is a convention introduced to make contact with Eq.(\ref{CHmetr}) below.

 The Atiyah--Hitchin metric  can also be written explicitly it terms of elliptic
functions \cite{AH}. It differs from (\ref{TAUBNUT}) by a series of exponentially
 suppressed at large distances terms. This series corresponds to a sum over
instantons in the 3d SYM theory.

The asymptotical TAUB-NUT metric (\ref{TAUBNUT}) is singular at $r= (2\pi)/g^2$.
 But the full Atiyah--Hitchin metric is not: there is no physical reason for the
kinetic energy to become singular when the distance between monopoles
becomes small or vanishes. Actually, the AH metric can be {\it reconstructed} from
the requirement that it describes a smooth hyper--K\"ahlerian manifold and concides
with Eq.(\ref{TAUBNUT}) in the asymptotics.

These results were generalized in Ref.\cite{CH} to $SU(N)$
groups. Again, the effective Lagrangian represents a
hyper--K\"ahlerian $\sigma$ model on the generalized
Atiyah--Hitchin manifold of dimension $4(N-1)$. This also
describes the dynamics of $N$ BPS monopoles \cite{GM}. In the asymptotics, one obtains a generalized TAUB-NUT metric
  \begin{eqnarray}
  \label{CHmetr}
 ds^2\ = \ A_{ml} d\vec{r}_m d\vec{r}_l + A^{-1}_{ml} \Lambda_m \Lambda_l\ ,
\end{eqnarray}
where  $A$ is the following $N\times N$ matrix:
  \be
  \label{Aml}
 A_{mm} &=& 1 - \frac {g^2}{4\pi} \sum_{l\neq m} \frac 1{|\vec{r}_m - \vec{r}_l|}\ \ \ \ \ \ \ \ \ \ \ \ \ \ ({\rm no\ summation\ over } \ m), \nonumber \\
A_{ml}  &=& \frac {g^2}{4\pi |\vec{r}_m - \vec{r}_l|},\
\ \ \ \ \ \ \ \ \ \ \ \ \ \ \ \ \ \ \ \ \ \ \ \ \ \ (m \neq l)\ ,
 \ee
 and
 $$ \Lambda_m = d\Psi_m +  \frac {g^2}{4\pi} \sum_{l \neq m}
\omega(\vec{r}_m - \vec{r}_l) \ .$$
 The explicit expressions for the generalized AH metric were not
found yet, but a {\it conjecture}
 of existence and uniqueness can be
 formulated:
there is only one smooth
hyper-K\"ahlerian manifold of dimension $4(N-1)$
 with the asymptotics (\ref{CHmetr}).

One can also consider the 2d and 1d dimensionally reduced versions
of  the original theory and study the corresponding effective
Lagrangians there. In the 2d case, this was done for $SU(2)$ in
Ref.\cite{DS} and for an arbitrary gauge group in Ref.\cite{Smilga}.
The effective Lagrangian also represents in this case a
supersymmetric $\sigma$ model on a manifold of $4r$ dimensions, where $r$ is the
rank of the group.
The coordinates of  target space correspond to four
 components of Abelian (belonging to the Cartan subalgebra) 6d vector potentials
associated with the reduced dimensions. In 2 dimensions, there are no dynamical
degrees of freedom associated with gauge fields.
The model involves four complex supercharges.
However, it is not a hyper--K\"ahlerian $\sigma$ model, but rather a
{\it twisted}  $\sigma$ model of the class described in Ref.\cite{GHR}.
In the simplest $SU(2)$ case, the bosonic part of the Euclidean Lagrangian takes
the form
\footnote{This expression differs from Eq.(3.23) of Ref.\cite{Smilga} by the presence of
the factor $i$, which appears after Wick rotation.}
   \be
  \label{Lbos2d}
 {\cal L}^{\rm bos}  \ =\
 \left[ 1 - \frac {g^2_2}{2\pi(\bar\sigma \sigma + \bar\phi \phi)} \right]
  \left[ \left| \partial_\mu \phi \right|^2 + \left| \partial_\mu \sigma \right|^2 \right] \  \nonumber \\
 + \frac {ig^2_2 \epsilon_{\mu\nu}}{2\pi(\bar\sigma \sigma + \bar\phi \phi)}
 \left[ \frac
 {\sigma}{\bar\phi}  (\partial_\mu \bar\sigma)(\partial_\nu \bar\phi)
+ \frac  {\bar\sigma} \phi
(\partial_\mu \sigma)(\partial_\nu \phi)
 \right]  \ ,
    \ee
 where $g_2^2$ is the 2d gauge coupling constant,
$\sigma$ is expressed into reduced Abelian gauge potentials of the original 4d theory, and
$\phi$ is the Abelian complex scalar
field there (if lifting up to  $d=6$,
$\phi$ is also expressed into 5-th and 6-th components
 of the gauge potentials).
 The first term in Eq.(\ref{Lbos2d}) describes a certain complex
metric (which is {\it not} K\"ahlerian).
The second twisted term can be associated with torsion.

The result (\ref{Lbos2d}) can be obtained by evaluating one--loop diagrams.
Higher loops vanish due to supersymmetry. And not only that.  In two
dimensions, ${\cal N} = 4$ supersymmetry together with rotational 
$O(4)$ invariance of the moduli space
rigidly fix the functional form of the metric, 
which makes the result
(\ref{Lbos2d}) {\it exact}. Another way to see this is to notice that
our 2d gauge theory (in contrast to its 3d and 4d "parents") does not
involve instantons of the usual type and the effective Lagrangian does not 
acquire
any nonperturbative contributions. 
\footnote{One can note, however,  that 
 2d non-Abelian gauge theories containing only adjoint fields
involve the so called $Z_N$ instantons \cite{ZN}.  
It would be interesting to pinpoint the precise reason by which
they do not contribute in this case.} 

A similar problem can be posed and solved for the quantum mechanical model
obtained after reduction of the original theory down to $(0+1)$ dimensions.
This was done in Ref.\cite{Smilga}. The effective
Lagrangian involves $5r$ bosonic variables.
It represents a generalization of the
nonstandard ${\cal N} = 4$ $\sigma$  model living on 5--dimensional
target space, which was constructed in Ref.\cite{DE}.

The only problem that has not been  solved yet is  constructing  the effective Lagrangians
for 3d theories with nonunitary gauge groups.
This is done in the present article. The basic observation which
allows one to obtain the results in a simple and universal way is that effective Lagrangians
in different dimensions are all related to each other.  The relationship between 4d and 3d
effective Lagrangians was
discussed back in \cite{WS2}. In Refs.\cite{Akhmedov,Smilga}, a similar  relationship between
1d and 2d effective Lagrangians was exploited to determine the form of the latter.  In this
 paper we start from the 2d effective Lagrangian and reconstruct the 3d one.

In the next section we illustrate our method by rederiving the results for $SU(2)$ by our
method. The generalization to all other groups is
rather straightforward. In Sect. 3 we explore in details the groups $Sp(2r)$, $SO(N)$ for odd
and even $N$, and $G_2$ and establish the
relationship of the corresponding hyper-K\"ahler manifolds with monopole moduli spaces.
The former are obtained  from the latter by the procedure of
{\it hyper-K\"ahlerian reduction}. Basically, it consists in our case
in imposing certain constraints on monopole dynamic variables which are compatible
with equations of motion.

In Sect. 4, we give a natural  D-brane
interpretation of the results obtained along the lines of \cite{HW} with insertion of the proper
orientifolds \cite{H}.

\section{Construction of ${\cal L}_{\rm eff}^{d=3}$}

 The basic idea is to consider the original  SYM theory not on $R^3$ and not  on $R^2$, but
rather on $R^2 \times S^1$, in the spirit of \cite{WS2,DS,Akhmedov,Smilga}. Playing with the
length $L$  of the circle, one can interpolate between 2d and 3d pictures.

The Lagrangian (\ref{Lbos2d}) was obtained after integrating out the charged fields in 2d theory. Thinking in 2d terms, we have now an infinite number of charged fields representing the coefficients in the Fourier series
  \be
  \label{Fourier}
  f(x,y;\ z) \ =\ \sum_{k = -\infty}^\infty  f_k(x,y) e^{2\pi i kz/L} \ .
  \ee
 The relevant variables in the effective Lagrangian are still
 zero Fourier modes of the neutral fields $\phi$, $\sigma$ and their superpartners.
The expression (\ref{Lbos2d}) is replaced by the infinite sum
 \be
 \label{LR2S1}
{\cal L}  \ =\
 \left[ 1 - \frac {g_2^2}{2\pi}  \sum_{n = -\infty}^\infty
\frac 1{(\bar\sigma_n \sigma_n + \bar\phi \phi)} \right]
  \left[ \left| \partial_\mu \phi \right|^2 + \left| \partial_\mu \sigma \right|^2
\right] \  \nonumber \\
 + i \sum_{n = -\infty}^\infty
 \frac {g_2^2\epsilon_{\mu\nu}}{2\pi(\bar\sigma_n
\sigma_n + \bar\phi \phi)}
 \left[ \frac
 {\sigma_n}{\bar\phi}  (\partial_\mu \bar\sigma)(\partial_\nu \bar\phi)  +\frac
 {\bar\sigma_n} \phi
(\partial_\mu \sigma)(\partial_\nu \phi)
 \right]  \ ,
\ee
where $\sigma_n = [Z + i(\tau + 2\pi n/L)]/\sqrt{2}$ (and also
$\phi = (X + iY)/\sqrt{2}$).
\footnote{The fields $X,Y,Z$ should not be confused with the spatial coordinates
$x,y,z$.}

In the limit $L \to 0$, only one term in the sum survives and we are reproducing the
 previous result. But for large $L \gg g_2^{-1}$ all terms are essential.
In the limit
 $L \to \infty$, we can actually replace the sum by an integral.  We obtain
 \be
 \label{Ld3dodual}
{\cal L} \ =\ \left(\frac 12 - \frac {g^2}{4\pi |\vec{r}|} \right)
\left[ (\partial_\mu \vec{r} )^2 + (\partial_\mu \tau )^2 \right] \nonumber \\
+ \frac {ig^2}{2\pi} \vecg{\omega}(\vec{r}) \epsilon_{\mu\nu} \partial_\mu \tau \partial_\nu \vec{r}\ ,
 \ee
where $g^2 = g_2^2 L$ is the 3--dimensional gauge coupling constant, $\vec{r} =
(X,Y,Z)$, and  $\omega(\vec{r}) = \vecg{\omega}(\vec{r}) d\vec{r}$ is
defined in Eq.(\ref{Dirac}).
 The variables $\vec{r}$ live on $R^3$ whereas the variable $\tau$ lives on the dual circle,
$0 \leq \tau \leq 2\pi/L$. When $L$ is very large, the size of the dual circle is very small
which would normally imply that the excitations related to nonzero Fourier modes of $\tau$
would become heavy and decouple. This happens, for example, when the 2d effective Lagrangian
is reconstructed with this method from
the 1d one \cite{Smilga}: the latter involves $5r$ dynamic bosonic degrees of freedom, while
the former --- only $4r$. But in our case it would not be correct just to cross out the terms
involving $\tau$. The presence of the twisted term $\propto \epsilon_{\mu\nu} $ prevents us to
do it.

To understand it, consider a trivial toy model,
 \be
 \label{toy}
 {\cal L} = \frac 12 (\dot{x}^2 + \dot{y}^2) + B x \dot{y} \ \ \Longrightarrow  \ \ H = \frac 12 \left[ p_x^2  + (p_y - Bx)^2 \right]\ ,
 \ee
where $x \in R^1$, while $y$ is restricted to lie on a small circle,
$0 \leq y \leq \alpha$. The Lagrangian (\ref{toy}) describes a particle living on a   cylinder and moving in a  constant magnetic field . Now, if the magnetic field $B$ were absent, the higher Fourier modes of the variable $y$ would be heavy and the low--energy spectrum would be continuous
corresponding to free motion along $x$ direction.
When $B \neq 0$, for {\it each} Fourier mode of the variable $y$, we obtain
 the {\it same} oscillatorial spectrum. Only the position of the center of the orbit and not the energy depends on $p_y^{(n)} = 2\pi n/\alpha$.

Thus, we cannot  suppress the variable $y$ in the Lagrangian (\ref{toy}). Likewise,  we cannot  suppress the variable $\tau$ in Eq.(\ref{Ld3dodual}). What we can do, however, is to trade it to another variable using the {\it duality} trick \cite{GHR,HKLR,duality}. Let us  write instead of (\ref{Ld3dodual}) another Lagrangian
 \be
 \label{promezh}
 {\cal L} \ =\ \left(\frac 12 - \frac {g^2}{4\pi |\vec{r}|} \right)
\left[ (\partial_\mu \vec{r} )^2 +  B_\mu^2 \right] \nonumber \\
+ i \epsilon_{\mu\nu} B_\mu  \left[ \partial_\nu \Psi +
\frac {g^2}{2\pi}
\vecg{\omega}(\vec{r})  \partial_\nu \vec{r} \right]\ .
 \ee
 Now, integrating $e^{-S_E}$ over $\prod d\Psi$ gives us $\epsilon_{\mu\nu} \partial_\nu
B_\mu = 0$ which implies that $B_\mu = \partial_\mu \tau$. Substituting it in Eq.(\ref{promezh}), we reproduce the result
(\ref{Ld3dodual}).
On the other hand, we can integrate over $\prod dB_\mu$ in Eq. (\ref{promezh}). Doing this,  we obtain the $\sigma$ model Lagrangian on the
manifold (\ref{TAUBNUT}) !
\footnote{A general statement of Ref.\cite{HKLR} is that, if the manifold corresponding to a twisted $\sigma$ model involves an isometry (the metric and torsion do not depend on some variables), the duality transfromations with respect to these variables brings us
onto a hyper--K\"ahlerian manifold. See Appendix for some further clarifications.}

This derivation can be readily generalized for other groups. The effective 2d Lagrangian for an arbitrary group depends on the variables $\vec{r}^a$, $\tau^a$ , $a = 1,\ldots,r$.
Putting the theory on $R^2 \times S^1$ and doing the sum over all Fourier harmonics
of charged fields (which for large $L$ can be replaced by the integral
  over $\prod_a d\tau^a$, we obtain the bosonic Euclidean
effective Lagrangian in the following form
  \be
 \label{Ld3dodual-r}
{\cal L} \ =\ \sum_j\left(\frac 1{c_V} - \frac {g^2}{4\pi |\vec{r}^{(j)}|} \right)
\left[ (\partial_\mu \vec{r}^{(j)} )^2 + (\partial_\mu \tau^{(j)} )^2 \right] \nonumber \\
+ \frac {ig^2}{2\pi}
\sum_j \vecg{\omega}(\vec{r}^{(j)}) \epsilon_{\mu\nu} \partial_\mu \tau^{(j)} \partial_\nu \vec{r}^{(j)}\ ,
 \ee
 where $\vec{r}^{(j)} = \alpha_j(\vec{r}^a)$, $\tau^{(j)} = \alpha_j(\tau^a)$, and the sum runs over all positive roots $\alpha_j$ of the corresponding Lie algebra. Performing the duality transformation and trading $\tau^a \to \Psi^a$, we obtain
  \be
  \label{L3d}
  {\cal L}\ =\ \frac 12 (\partial_\mu \vec{r}^a )  (\partial_\mu \vec{r}^b ) Q_{ab} + \frac 12  J_\mu^a Q^{-1}_{ab} J_\mu^b \ ,
  \ee
  where
  \be
  \label{QiJ}
 Q^{ab} &=& \delta^{ab} - \frac {g^2}{2\pi} \sum_j \frac
 {\alpha_j^a \alpha_j^b}{|\vec{r}^{(j)}|}\ , \nonumber \\
 J_\mu^a &=&  \partial_\mu \Psi^a +
 \frac {g^2}{2\pi} \sum_j \vecg{\omega}(\vec{r}^{(j)})
 \partial_\mu \vec{r}^{(j)} \alpha_j^a \ .
  \ee
The relation $\sum_j \alpha_j^a \alpha_j^b
\ =\ (c_V/2) \delta^{ab}$ (with the normalisation
$\sum_a \alpha_j^a \alpha_j^a = 1$ for the long roots) was
used.

\section{${\cal L}_{\rm eff}$ and monopole dynamics.}

Specialization of Eq.(\ref{L3d}) to particular groups is
naturally interpreted
 in terms of metrics on BPS monopoles moduli spaces.

{\it i)} $SU(N)$. This case was first considered in \cite{CH}. There are $N(N-1)/2$
positive roots, $\alpha_{ml}(\vec{r}) = \vec{r}_m - \vec{r}_l,
m < l = 1,\ldots,N,\ \sum_m \vec{r}_m = 0$.
Substituting
 it in Eq. (\ref{L3d}), we reproduce the result
(\ref{CHmetr}).
\footnote{   Note that for  simply laced $SU(N)$, there is no difference
between roots and coroots.
There is such difference  - and this will be important below -
in $Sp(2r)$ and $SO(2r+1)$ cases. }

The monopole dynamics is described by the  following classical
equations of motion \cite{GM}
\footnote{Here $g$ is interpreted as the monopole magnetic charge. In the quantum problem, $q_l$ are quantized to (integer)/$g$ and are interpreted as the electric charges of the corresponding dyons.}
 \be
 \label{eqmotmon}
 \ddot{\vec{r}}_l - \frac {g^2}{4\pi} \sum_{m\neq l}^N \frac {\ddot{\vec{r}}_{lm} }{\vec{r}_{ml}} + \frac {g^2}{8\pi} \sum_{l \neq m = 1}^N \frac{2\left[\dot{\vec{r}}_{ml} \times \vec{r}_{ml}
\right] \cdot \dot{\vec{r}}_{ml} - \vec{r}_{ml} (\dot{\vec{r}}_{ml}^2) }{r_{ml}^3} \nonumber \\
- \frac {g}{4\pi} \sum_{m \neq l  } (q_{ml}) \dot{\vec{r}}_{ml} \times \frac {\vec{r}_{ml}}{r_{ml}^3} + \frac 1{8\pi} \sum_{m \neq l} \frac { q_{ml}^2 \vec{r}_{ml}}{r_{ml}^3} \ = 0\ , \nonumber \\
q_l \ =\ gA^{-1}_{lm} \left[ \dot{\Psi}_m + \frac {g^2}{4\pi}
\sum_{n \neq m} \vecg{\omega}( \vec{r}_{nm} ) \dot{\vec{r}}_{nm} \right] \ =\ {\rm const}\ ,
 \ee
where $\vec{r}_{nm} = \vec{r}_m - \vec{r}_l,\ q_{ml} = q_m - q_l$.

{\it ii)} $Sp(2r)$. There are $r$ long positive
roots $\alpha_m (\vec{r}) =  \vec{r}_m $
and $r(r-1)$
short positive roots   $\alpha_{ml}(\vec{r}) =
(\vec{r}_m \pm \vec{r}_l)/2$ ($m < l = 1,\ldots, r$ ; $\vec{r}_m$ are mutually orthogonal and linearly independent). The
metric reads
\begin{eqnarray}
\label{C}
ds^2\ =\ \sum_m (d{\bf r}_m)^2 - \frac{g^2}{4\pi} \sum_{\pm}
\sum_{m<l}
\frac{(d {\bf r}_l \pm d{\bf r}_m)^2}
 {\vert{\bf r}_l \pm {\bf r}_m \vert}
-   \frac{g^2}{2\pi}  \sum_m \frac{(d{\bf r}_m)^2}{r_m} +\ {\rm phase\ part} \nonumber \\
 \equiv
Q_{ml} d{\bf r}_m d{\bf r}_l +\ {\rm phase\ part}\ .
\end{eqnarray}
The full metric is restored from Eqs.(\ref{L3d}, \ref{QiJ}).

An important observation is that the
corresponding effective Lagrangian (the QM version thereof)
is obtained from  the effective Lagrangian describing the dynamics
 of $2r+1$ BPS monopoles
numbered by the integers $j=-r,\ldots, r$
by imposing the constraints
\be
\label{A2C}
\vec{r}_{-r} + \vec{r}_{r} \ = \dots = \vec{r}_{-1} + \vec{r}_{1} \ =\ 2\vec{r}_0 \ =\ 0 \ , \nonumber \\
\Psi_{-r} + \Psi_{r} \ = \dots = \Psi_{-1} + \Psi_{1} \ =\ 2\Psi_0 \ =\ 0\ .
 \ee
 We {\it are} allowed to impose these constraints because they are compatible with the equations of motion (\ref{eqmotmon}) and also with the equations of
motion of the corresponding 2d field theory.

The corresponding  metric is hyper-K\"ahlerian. It follows from: 
{\it (i)} ${\cal N} = 4$ 
supersymmetry of the original theory, which implies 
${\cal N} = 4$ supersymmetry of the effective Lagrangian (\ref{L3d}),
{\it (ii)} the absence of the twisted term there, and  {\it (iii)}
the theorem due to Alvarez--Gaum\'e and Freedman \cite{AGF}.
One can also demostrate the hyper-K\"ahlerian nature of the metric more 
directly by reproducing the result (\ref{B})
in the framework of the  hyper-K\"ahlerian reduction procedure described 
in \cite{GR}. 
The reduction of the Gibbons-Manton metric (\ref{CHmetr}) is performed with 
respect   to the symmetry
\begin{equation}
\label{s1}
\Psi_j \rightarrow \Psi_j + \alpha_j,\; \Psi_{-j} \rightarrow \Psi_{-j} + \alpha_j, j=1, \ldots, r,
\Psi_0  \rightarrow \Psi_0 + \alpha_0,
\end{equation}
where $\Psi_j$ is the phase variable of the $j$th monopole.
The corresponding moment maps are
\begin{equation}
\label{m1}
 {\bf r}_0, {\bf r}_j+{\bf r}_{-j},\; j=1,\ldots,r,
\end{equation}
and  the hyper--K\"ahler reduction is made at zero value of the moment maps.

{\it iii)} $SO(2r+1)$. The system of roots is the same as for $Sp(2r)$ only the long and short roots are interchanged: there are now $r(r-1)$ long roots
$(\vec{r}_m \pm \vec{r}_l)/\sqrt{2}$ and $r$ short roots $\vec{r}_m/\sqrt{2}$.
 The metric reads
\begin{eqnarray}
\label{B}
ds^2\ =\
\sum_m (d{\bf r}_m)^2 - \frac{g^2}{2\pi \sqrt{2}} \left[ \sum_{\pm}
\sum_{m<l}
\frac{(d {\bf r}_l \pm d{\bf r}_m)^2}
 {\vert{\bf r}_l \pm {\bf r}_m \vert} + \sum_m \frac{(d{\bf r}_m)^2}{r_m}
\right] \nonumber \\
+\ {\rm phase\ part}\ .
\end{eqnarray}
This metric is obtained from the Gibbons-Manton type  metric for $2r$ BPS monopoles
numbered by the integers $j=-r, \ldots, r,\; j\neq 0$ by
imposing the constraints
  \be
\label{A2B}
\vec{r}_{-r} + \vec{r}_{r} \ = \dots = \vec{r}_{-1} + \vec{r}_{1} \ =\ 0 \ , \nonumber \\
\Psi_{-r} + \Psi_{r} \ = \dots = \Psi_{-1} + \Psi_{1} \  =\ 0\
 \ee
 and rescaling $ds^2$ and $g^2$.
The constraints (\ref{A2B}) are compatible with the equations of motion. The result (\ref{B}) is also obtained by
 hyper--K\"ahler reduction with
respect to the symmetry
\begin{equation}
\label{s2}
\Psi_j \rightarrow \Psi_j + \alpha_j,\; \Psi_{-j} \rightarrow \Psi_{-j} + \alpha_j, j=1, \ldots, r\ .
\end{equation}
The corresponding moment maps are
\begin{equation}
\label{m2}
{\bf r}_j+{\bf r}_{-j},\; j=1,\ldots,r,
\end{equation}
and the hyper-K\"ahler reduction is made at zero value of the moment maps.

Note that we obtained the effective Lagrangian for $Sp(2r)$ out of that for $SU(2r+1)$ and not out of
$SU(2r)$, as one could naively expect in view of the embedding
$Sp(2r) \subset SU(2r)$. Likewise, the moduli space for $SO(2r+1)$ is
obtained out of  $SU(2r)$  and not  $SU(2r+1)$. This is  due to
the fact that
magnetic charges are coupled to coroots rather than roots.

{\it iv)} $SO(2r)$. This Lie algebra is simply laced. The positive roots are
$\alpha_{ml}^\pm(\vec{r}) = (\vec{r}_m \pm \vec{r}_l)/\sqrt{2}$
($m < l = 1,\ldots r$)

The metric reads
\begin{eqnarray}
\label{D}
ds^2\ =\
\sum_m (d{\bf r}_m)^2 - \frac{g^2}{2\pi \sqrt{2}}  \sum_{\pm}
\sum_{m<l}
\frac{(d {\bf r}_l \pm d{\bf r}_m)^2}
 {\vert{\bf r}_l \pm {\bf r}_m \vert} \ + \ldots
\end{eqnarray}
To relate this metric to the Gibbons-Manton
 one, we need first to introduce a massive deformation of the latter,
\begin{eqnarray}
\label{mass}
 ds^2\ =\
\sum_m (d{\bf r}_m)^2 -  \frac{g^2}{4\pi} \sum_{l < m}
\frac {(d{\bf r}_m - d{\bf r}_l)^2}{\sqrt{({\bf r}_m - {\bf r}_l)^2 +
\lambda_{lm}^2}} + \ldots\ ,
\end{eqnarray}
where there are $2r$ monopoles numbered by the integers
$j=-r, \ldots, r,\; j\neq 0$.
This metric is hyper--K\"ahler \cite{GR}.
Assuming that only $\lambda_{m,-m}$ are not zero, sending these parameters to infinity, and
performing the hyper--K\"ahler reduction with respect to the symmetry
\begin{equation}
\label{s3}
\Psi_j \rightarrow \Psi_j + \alpha_j,\; \Psi_{-j} \rightarrow \Psi_{-j} + \alpha_j, j=1, \ldots, r,
\end{equation}
with moment maps
\begin{equation}
\label{m3}
{\bf r}_j+{\bf r}_{-j},\; j=1,\ldots,r,
\end{equation}
and at zero value of the moment maps,  Eq.(\ref{mass}) is reduced to  Eq.(\ref{D}) after a proper rescaling.

{\it v)} $G_2$ There are three long ( $\vec{r}_1 - \vec{r}_2, \vec{r}_1 - \vec{r}_3, \vec{r}_2 - \vec{r}_3$) and three short ($\vec{r}_{1,2,3}$)
positive roots (the constraint $\vec{r}_1 + \vec{r}_2 + \vec{r}_3 =
 0 $  being imposed). The metric reads
 \be
 \label{G}
 ds^2 \ =\ \sum_{m=1}^3 d\vec{r}_m^2 - \frac {g^2}{2\pi}
 \left( \sum_{m>l=1}^3 \frac {(d\vec{r}_m  - d\vec{r}_l)^2}{|\vec{r}_m  - \vec{r}_l|}  + 3 \sum_{m=1}^3 \frac {d\vec{r}_m^2}{|\vec{r}_m |} \right) + \ldots\ .
  \ee
It can be obtained out of the metric for $Sp(6)$
  \be
 \label{Sp6}
 ds^2 \ =\ \sum_{m=1}^3 d\vec{r}_m^2 -
 \frac {g^2 }{4\pi}
 \left( \sum_\pm \sum_{m>l=1}^3 \frac {(d\vec{r}_m  \pm d\vec{r}_l)^2}{|\vec{r}_m  \pm \vec{r}_l |}  + 2 \sum_{m=1}^3 \frac {d\vec{r}_m^2}{|\vec{r}_m |} \right) + \ldots
  \ee
  by rescaling and  imposing the [compatible with $Sp(6)$ equations of motion] constraints
\be
  \label{C3-2-G2}
\vec{r}_1 + \vec{r}_2 + \vec{r}_3 \ =\ 0\ , \ \ \ \ \ \ \
\Psi_1 + \Psi_2 + \Psi_3 \ =\ 0\ .
  \ee
  Again, we obtained ${\cal L}_{\rm eff}$ for  $G_2$ out of   ${\cal L}_{\rm eff}$ for  $Sp(6)$, though    $G_2$ is embedded not into
$Sp(6)$, but into the dual algebra $SO(7)$.

The metric Eq.(\ref{G}) is obtained from the $Sp(6)$ metric by the  hyper-K\"ahler reduction with respect to symmetry
\be
\label{g2}
\Psi_1 \rightarrow \Psi_1 +\alpha, \; \Psi_2 \rightarrow \Psi_2 +\alpha, \; \Psi_3 \rightarrow \Psi_3 +\alpha
\ee
with moment map
\be
  \label{g22}
\vec{r}_1 + \vec{r}_2 + \vec{r}_3 \,
  \ee
at zero value of the moment map.

The effective Lagrangian for $F_4$  can be related to the moduli space of
26 monopoles. (26 is the lowest dimension of a unitary group where
$F_4$ can be embedded. This follows from the fact that the representation {\bf 26} of $F_4$ has the lowest dimension.) $E_6$ can be embedded into
$SU(27)$ and hence the corresponding effective Lagrangian is
 related to the moduli space of
27 monopoles. Now, the shortest representation in $E_7$ has the
dimension 56 and we need at least 56 monopoles in this
case. Finally, $E_8 \subset SU(248)$ and we need 248 monopoles. The moduli
space of 248 monopoles can also be used as a universal starting point
to describe the dynamics
of $F_4, E_6$ and $E_7$, if
following the chain of embeddings $F_4 \subset E_6 \subset E_7 \subset E_8
\subset SU(248)$.

The  explicit formulae we have written refer to the asymptotic region
 where nonperturbative effects are suppressed. The corresponding metrics
involve singularities at small $|\vec{r}^{(j)}|$.
Like for the $SU(N)$ case, a very reasonable conjecture is that  these
singularities can be sewn up and, for any simple Lie group, there is one
and only one smooth hyper-K\"ahler metric with the asymtotics
\be
ds^2 \ =\ d\vec{r}^a Q_{ab} d\vec{r}^b + \ldots
 \ee

It is  natural to conjecture that this metric is obtained
from the multimonopole Atiyah-Hitchin
metrics by the hyper-K\"alerian reduction with respect to the same symmetries
as above, Eqs. (\ref{s1}),(\ref{s2}),(\ref{s3}),(\ref{g2}).

\section{Orientifolds.}

We now discuss the brane pictures behind the results obtained above.

Let us first remind \cite{HW} that (the Coulomb branch of)
the ${\cal N}=4$ 3d SUSY Yang-Mills with gauge group
$U(N)$ \footnote{The common $U(1)$ is standard in the brane pictures and is
essentially irrelevant.}
is convenient to realize
as $N$ parallel $D3$ branes stretched between two parallel $NS5$ branes.
One of the directions on $D3$ brane
is of microscopic size (the distance between  $NS5$ branes) so from the $D3$ branes perspective there is
$3d$ gauge theory. Low energy degrees of freedom are positions of the $D3$ branes on the $NS5$ branes
which gives $3N$ scalars (the branes assumed to be solid in this counting) and $N$ photons living
on the branes, which gives other $N$ scalars. Forgetting about the common $U(1)$ leaves $3(N-1)$ scalar degrees
of freedom.
The charged particles corresponding to the roots of $U(N)$ appears as $F$ string states
corresponding to the strings stretched between two of the $N$ $D3$ branes, in particular, simple roots
correspond to the strings stretched between two adjacent branes of the $N$ $D3$ branes, {\it elementary} strings.
Attributing
to the $j$th brane the element $e_j$ of the orthonormalized Euclidean basis $\{e_j,\;j=1,\ldots,n \}$ one sees
that the elementary F strings inherit vectors of the type of $e_{j+1}-e_j$. It is easy to see that intersections
of these vectors are described by $SU(N)$ Dynkin diagrams. One can also check that there is the appropriate amount
of supersymmetry in the brane system described.

As explained in \cite{HW}, with use of the results of 
\cite{Diaconescu}, the $SU(2)$ monopole moduli space appears
if one changes the perspective from $D3$ branes to 
$NS5$ branes.  Two parallel $NS5$ branes give
$U(2)$ gauge theory, positions of the $N$ $D3$ branes 
on the $NS5$ branes give positions of $N$ monopoles
and the $N$ scalars dual to photons living on the $D3$ branes 
give $N$ phases to the monopoles.
Forgetting about the center-of-mass coordinate and about the 
common phase gives $3(N-1)$ dimensional
relative moduli space of $SU(2)$ monopoles.

Consider now the case of $Sp(r)$ and of $SO(2r+1)$.

In the $Sp(r)$ case the brane picture consists of $2r$ $D3$ branes and $O^+$  orientifold
in the middle \cite{H}. Let us number the branes by integers $j=-r, \ldots, r,\; j \neq 0$.
The locations of $D3$ branes are symmetric with respect to orientifold, ${\bf r}_j=-{\bf r}_{-j}$.
Again, one attributes to the $j$th brane ($j>0$)
the element $e_j$ of the orthonormalized Euclidean basis $\{e_j,\;j=1,\ldots,r \}$ and
one defines $e_{-j}=-e_j$. The elementary $F$ strings are of two types - those crossing the orientifold and those
not crossing the orientifold. The ones not crossing the orientifold appear only in symmetric pairs under the reflection
with respect to the orientifold, while the ones crossing orientifold are required to be self-dual under the reflection.
Thus, the elementary strings inherit vectors $2e_1, e_2-e_1, \ldots, e_j-e_{j-1}$, whose intersections are described by
$Sp(r)$ Dynkin diagrams.

In the $SO(2r+1)$ case the brane picture consists of $2r$ $D3$ branes and ${\tilde O}^-$ orientifold.
Everything is quite similar to the case of $Sp(r)$ but the rules for elementary $F$ strings are  different.
They are  of two types - those ending on the orientifold and those not ending on the orientifold.
Crossing of the orientifold is not now allowed. All the strings appear only in symmetric pairs under the reflection
with respect to the orientifold. The elementary strings this time inherit vectors $e_1, e_2-e_1, \ldots, e_j-e_{j-1}$,
whose intersections are described by $SO(2r+1)$ Dynkin diagrams.

Magnetically charged states appear as $D$ strings stretched between the $D3$ branes. The orientifolds introduce different rules
for $F$ strings and $D$ strings. Actually, the rules for $D$ strings in the presence of $O^+$ orientifold
are the same as for $F$ strings in the presence of ${\tilde O}^-$ and vice versa. This fits nicely with the metrics Eqs.(\ref{B}), (\ref{C})
in the cases of $Sp(r)$ and $SO(2r+1)$.

In the $SO(2r)$ case the brane picture consists of $2r$ $D3$ branes and 
${O}^-$ orientifold.
The elementary $F$ (and $D$) strings can now cross the orientifold but cannot be self-dual
under the reflection with respect to the orientifold. They thus inherit the vectors  $e_2+e_1, e_2-e_1, \ldots, e_j-e_{j-1}$
whose intersections are described by $SO(2r)$ Dynkin diagrams.
This is the limit $\lambda \rightarrow \infty$ which removes from metric Eq.(\ref{D}) the terms which would be present
if the self-dual strings were  present.

In the case of  $G_2$ the orientifold picture is not known. Considerations of the previous section (cf. Eqs.(\ref{g2}), (\ref{g22})) suggest to conjecture
that a novel kind of  orientifold relating positions of three branes (not of two ones like in $SO/Sp$ case) is  necessary.

\section{Aknowledgements.}

We are indebted to N. Hitchin, V. Kac, and  V. Rubtsov for useful discussions
and correspondence. K.S. acknowledges kind hospitality at the SUBATECH (Nantes), where this work was done.

\section*{Appendix: Twisted, Hyper-K\"ahler, and Duality.}

To make the paper more self--contained, we present here some relevant formulae referring to manifestly supersymmetric description of the twisted $\sigma$ models and the duality transformation relating them to hyper--K\"ahlerian models.

 The Lagrangian of a twisted supersymmetric $\sigma$ model
generically reads
\begin{equation}
\label{Ltwist}
L=\int d^2\theta d^2{\bar \theta}\ {\cal K}_{tw}(\Phi^{a}, {\bar \Phi}^{a};\   \Sigma^{a}, {\bar \Sigma}^{a})
\end{equation}
 where $\Phi^a$ is a ${\cal N}=2$ chiral superfield,
\begin{equation}
{\bar D}_+\Phi^a={\bar D}_-\Phi^a=0,
\end{equation}
and $\Sigma^a$ is a  ${\cal N}=2$ twisted chiral superfield,
\begin{equation}
{D}_+\Sigma^a={\bar D}_-\Sigma^a=0\ .
\end{equation}

The Lagrangian (\ref{Ltwist}) is manifestly ${\cal N}=2$  supersymmetric. The requirement of
${\cal N}=4$ supersymmetry imposes the constraints
\begin{equation}
\label{constraint1}
\frac{{\partial}^2}{\partial \Sigma^{a}\partial {\bar \Sigma}^b}{\cal K}_{tw}-
\frac{{\partial}^2}{\partial \Sigma^{b}\partial
{\bar \Sigma}^{a}}{\cal K}_{tw}=0
\end{equation}
\begin{equation}
\label{constraint2}
\frac{{\partial}^2}{\partial \Phi^{a}\partial {\bar \Phi}^{b}}{\cal K}_{tw}+
\frac{{\partial}^2}{\partial \Sigma^{b}\partial {\bar \Sigma}^a}{\cal K}_{tw}=0.
\end{equation}

The target space metric is
\be
\label{metric}
ds^2 \ = \
\frac{\partial^2 {\cal K}_{tw}}   {\partial \Phi^a \partial {\bar \Phi}^b}
d \Phi^a d{\bar \Phi}^b \ +
\frac{\partial^2 {\cal K}_{tw}}
 {\partial \Sigma^a \partial {\bar \Sigma}^b}
 d \Sigma^a d {\bar \Sigma}^b \ .
\ee
We emphasize that in spite of ${\cal N}=4$ supersymmetry, the target space is not hyper-K\"ahler and not even K\"ahler. The
Lagrangian involves also a torsion term, which
 can be expressed via ${\cal K}_{tw}$ as well.

The twisted potential for the effective Lagrangian of 2d SYM theory is
\begin{eqnarray}
  \label{Kr}
   {\cal K}_{tw} \ =\ \sum_j \left\{ \frac {1}{2c_V}
\left[\bar\Sigma^{(j)} \Sigma^{(j)} - \bar\Phi^{(j)} \Phi^{(j)} \right] \right.
\nonumber \\
\left. -   \frac {g^2}{8\pi}\left[ F \left( \frac {\bar\Sigma^{(j)} \Sigma^{(j)} }
{\bar\Phi^{(j)} \Phi^{(j)}} \right)  - \ln \Phi^{(j)} \ln \bar\Phi^{(j)}
 \right]
\right\}
\ ,
 \end{eqnarray}
where
$ F(\eta)$ is the  Spence function,
\begin{equation}
   \label{Spence}
   F(\eta) \ =\ \int_1^\eta \frac {\ln(1+\xi)}\xi \ d\xi,
    \end{equation}
 Putting the SYM theory on $R^2 \times S^1$ and "unwinding" $S^1$, we obtain
another twisted $\sigma$ model, where
${\cal K}_{tw}$ depends only on the sums $Z^a =  \Sigma^a +
\bar \Sigma^a$ and does not depend on the differences.
It  satisfies the generalized harmonicity condition (\ref{constraint2}).( The constraint (\ref{constraint1}) is  satisfied automatically.)

In this case, Eq.(\ref{Ltwist}) can be rewritten as
   \begin{equation}
\label{promezhsusy}
L=\int d^2\theta d^2{\bar \theta}
\left[
{\cal K}_{tw} (\Phi^{a}, {\bar \Phi}^{a}, Z^a)
- \sum_a (A^a + \bar A^a)Z^a \right]\ ,
   \end{equation}
where $Z^a$ are real superfields and $A^a$ are conventional chiral
superfields. Integrating over $\prod_a dA^a d\bar A^a$, we  obtain
$D^2 Z^a = \bar D^2 Z^a = 0 $, and that implies that $Z^a$ can be
represented as $\Sigma^a + \bar \Sigma^a$, where $\Sigma^a$ are twisted
superfields. We are thus reproducing Eq.(\ref{Ltwist}).
On the  other hand, integrating out the superfields $Z^a$, we obtain
$$
\int d^2\theta d^2{\bar \theta}\
{\cal K}_{hk}(\Phi^{a}, {\bar \Phi}^{a}, X^a)\ ,
$$
where $X^a =
 A^a + {\bar A}^a $ and
${\cal K}_{hk}$ is related to ${\cal K}_{tw}$  by a
Legendre transformation
  \begin{equation}
\label{Legendre}
{\cal K}_{hk}({\Phi}^a, {\bar \Phi}^a, X^a) =
{\cal K}_{tw}({\Phi}^a, {\bar \Phi}^a, Z^a) - \sum_a Z^a X^a\ ,
   \end{equation}
where $Z^a$ as a function of $X^a$ is defined from the equation
   \begin{equation}
\label{X}
\frac{\partial {\cal K}_{tw}}{\partial Z^a} - X^a=0\ .
   \end{equation}
Now, ${\cal K}_{tw}$ satisfies {\it linear} generalized harmonicity
conditions (\ref{constraint2}). This dictates certain
{\it nonlinear} conditions on the hyper--K\"ahler potential
${\cal K}_{hk}$. For 4--dimensional manifolds, a harmonic
function ${\cal K}_{tw}$ determines the function
${\cal K}_{hk}$ satisfying the so called {\it Monge--Ampere}
equation
\be
\label{Monge}
\det \left \|
\begin{array}{cc}
\partial^2 {\cal K}_{\it hk}/
\partial {\bar \Phi} \partial   \Phi  &
\partial^2 {\cal K}_{\it hk}/\partial \bar \Phi \partial  X  \\
\partial^2 {\cal K}_{\it hk}/ \partial X \partial
 \Phi   &
\partial^2 {\cal K}_{\it hk}/\partial X^2
   \end{array} \right \| = {\rm const}
 \ee
The nonlinear equation (\ref{Monge}) is must more diffucult
to solve than the Laplace equation. If you will, solving
first the Laplace
equation for ${\cal K}_{tw}$ and applying the Legendre
transformation afterwards represents a regular method of finding  solutions to the Monge--Ampere equation
and, correspondingly, a wide class of hyper--K\"ahlerian
manifolds \cite{Roubtsov} (the functions
${\cal K}_{\it hk}$ not depending on $\bar A - A$ are thus found).
Note that in all nontrivial cases, the solution of the
differential
equation (\ref{X}) is  expressed into certain
trancendental
 functions, for which Ryzhik and Gradstein failed to reserve
 a symbol.

The hyper--K\"ahler metric is  computed via
the standard formulas
\begin{equation}
\label{metrichk}
g_{i {\bar j}}=\partial_i \partial_{\bar j} {\cal K}_{hk}.
\end{equation}
In contrast to ${\cal K}_{hk}$, it is often expressed into
conventional functions
[see e.g. Eqs.(\ref{TAUBNUT}), (\ref{CHmetr})].

\end{document}